# Low-Frequency Resonance in Strong Heterogeneity

Yinbin Liu
Burnaby, British Columbia, Canada
(Email: yinbinliu77@gmail.com)

**Abstract**
Multiple scattering of wave in strong heterogeneity can cause resonance-like wave anomaly where the signal exhibits low-frequency, high intensity, and slowly propagating wave packet velocity. For example, long period event in volcanic seismology and plasma oscillations in wave-particle interactions. Collective behaviour in a many-body system is thought to be the source for generating the anomaly, however the detailed mechanism is not fully understood. Here I show that the physical mechanism is associated with low-frequency resonance (LFR) in strong small-scale heterogeneity through seismic wave field modeling for bubble cloud heterogeneity and 1D heterogeneity. LFR is a kind of wave coherent scattering enhancement or emergence phenomenon. Its resonance frequency decreases with increasing heterogeneous scale, impedance contrast, or random heterogeneous scale and velocity fluctuations; its intensity diminishes with decreasing impedance contrast or increasing random heterogeneous scale and velocity fluctuations. LRF exhibits the characteristics of localized wave in space and the shape of ocean wave in time and is a ubiquitous wave phenomenon in wave physics. The concept of LFR can open up new opportunities in many aspects of science and engineering.

**Introduction**

A resonance appears when the frequency of a driving force matches the natural frequency of a system, which exhibits features of selective frequency and trapped energy. The wavelength of the resonance system is close to or smaller than the size of the system. The ringing of a bell is associated with this kind of wave phenomenon.

There is also a ubiquitous resonance-like wave phenomenon that can be observed in strong small-scale heterogeneity where multiple scattering of wave gives rise to a low-frequency anomaly with high intensity and slowly propagating wave packet velocity. Low frequency in this context means the dominant wavelength of signal is much larger than the heterogeneous scale of the system. For example, long period event in volcanic tremor (*1, 2*) and hydraulic fracturing microseismicity (*3*) in seismology and plasma oscillations (*4, 5*) and quantum Hall effects (*6, 7*) in wave-particle interactions. The collective behaviour or self-organization and synchronization of a many-body system is generally thought to be the source for generating the low-frequency anomalies, however the detailed physical mechanism is not well understood.

Strong small-scale (or microscopic) heterogeneity is a kind of complex many-body physics system that exhibits the nature of the hierarchical structure of science. The multiple scattering among many bodies can emerge an entirely new physical phenomenon that cannot be understood in terms of a simple extrapolation of system constituent units (*8*). Classical multiple wave scattering theory in a many-body system, based on wave equations and boundary conditions, provides a unified theoretical framework for understanding the origin of the macroscopic collective behaviour and revealing the physics of the microscopic constituent interactions. At present, we have not yet developed an effective algorithm, which includes all multiple scattering effects, to simulate wave scattering in a two-dimensional (2D) or three-dimensional (3D) many-body system. However, Foldy's multiple scattering theory may be employed to approximately study wave scattering in bubble cloud heterogeneity, Thomson-Haskell propagator matrix approach for a horizontally-layered system (*9*), which includes all multiple scattering, may be employed to study wave scattering in 1D heterogeneity. There exists a large number of studies on improving the numerical instability in the propagator matrix and simulating the wave propagation and scattering in weak 1D heterogeneity (impedance contrast < 5). This study will develop delta propagator matrix approach (*10*), which can effectively improve the numerical instability, to study the wave scattering in strong 1D heterogeneity (impedance contrast > 10). Based on scattered seismic wave field modelling for bubble cloud heterogeneity and 1D heterogeneity, I show that multiple scattering of wave in strong small-scale heterogeneity may excite low-frequency resonance (LFR) in transient regime. The concept of LFR provides a physical interpretation on the observed resonance-like wave phenomenon in strong small-scale heterogeneity

**Sommerfeld and Brillouin Precursors**

An electromagnetic pulse propagating through a single resonance Lorentz dielectric medium will be scattered into high-frequency Sommerfeld precursor and low-frequency Brillouin precursor (*11*). An acoustic pulse propagating through a bubble cloud medium may also exhibit wave packet evolution similar to Sommerfeld and Brillouin precursors. Fig. 1 shows the acoustic wave field (Figs. 1A to 1D), transmission coefficient (Fig. 1E), and normalized power spectrum (Fig. 1F) of the first cyclic low-frequency wave for acoustic wave scattering by gas-bearing magma medium (*12*, *13*) with different bubble radius and number (see Supplemental material). The other parameters are $\gamma = 1.1$, $\rho_f = 2{,}700$ kg/m$^3$, $v_f = 1{,}600$ m/s, $P_0 = 2.0 \times 10^5$ Pa, $z$ = 10 m and 100 m, $b = 0.01\omega_0$ ( $\omega_0 = 2\pi f_0$, $f_0$ is the Minnaert resonance frequency of a single bubble vibration). The principal branch or the first Riemann sheet ($-\pi \le arctg\left[\mathrm{Im}(k^2)/\mathrm{Re}(k^2)\right] \le \pi$ ) is chosen in numerical integration. It can be seen that the total field in Fig. 1A is composed of the early arrival high-frequency small-amplitude wave packet and the late arrival low-frequency large-amplitude wave packet. The former corresponds to Sommerfeld precursor and the latter corresponds to Brillouin precursor in a single resonance Lorentz dielectric medium (*11*). Sommerfeld precursor exhibits first exponentially increasing oscillation and then exponentially decaying oscillation, and its instantaneous frequency monotonically decreases from infinite (or the maximum frequency of source) to near the upper stopband corner frequency of the system (Fig. 1E). Brillouin precursor exhibits first monotonically increasing and then exponentially decaying oscillation, and its instantaneous frequency monotonically increases from zero (or the minimum frequency of source) to near the lower stopband corner frequency of the system (Fig. 1E). Brillouin precursor behaves as low-frequency, large-amplitude, and slowly propagating wave packet velocity. It exhibits the shape of ocean wave and can be described by the hyper-Airy function *(11)*. For short propagation distance, Sommerfeld and Brillouin precursor fields will partially overlap and show the feature of long period event that consists of a high-frequency small-amplitude onset superposing on a low-frequency large-amplitude background in volcanic tremor (*1*, *2*) and in hydraulic fracturing stimulation (*3*).

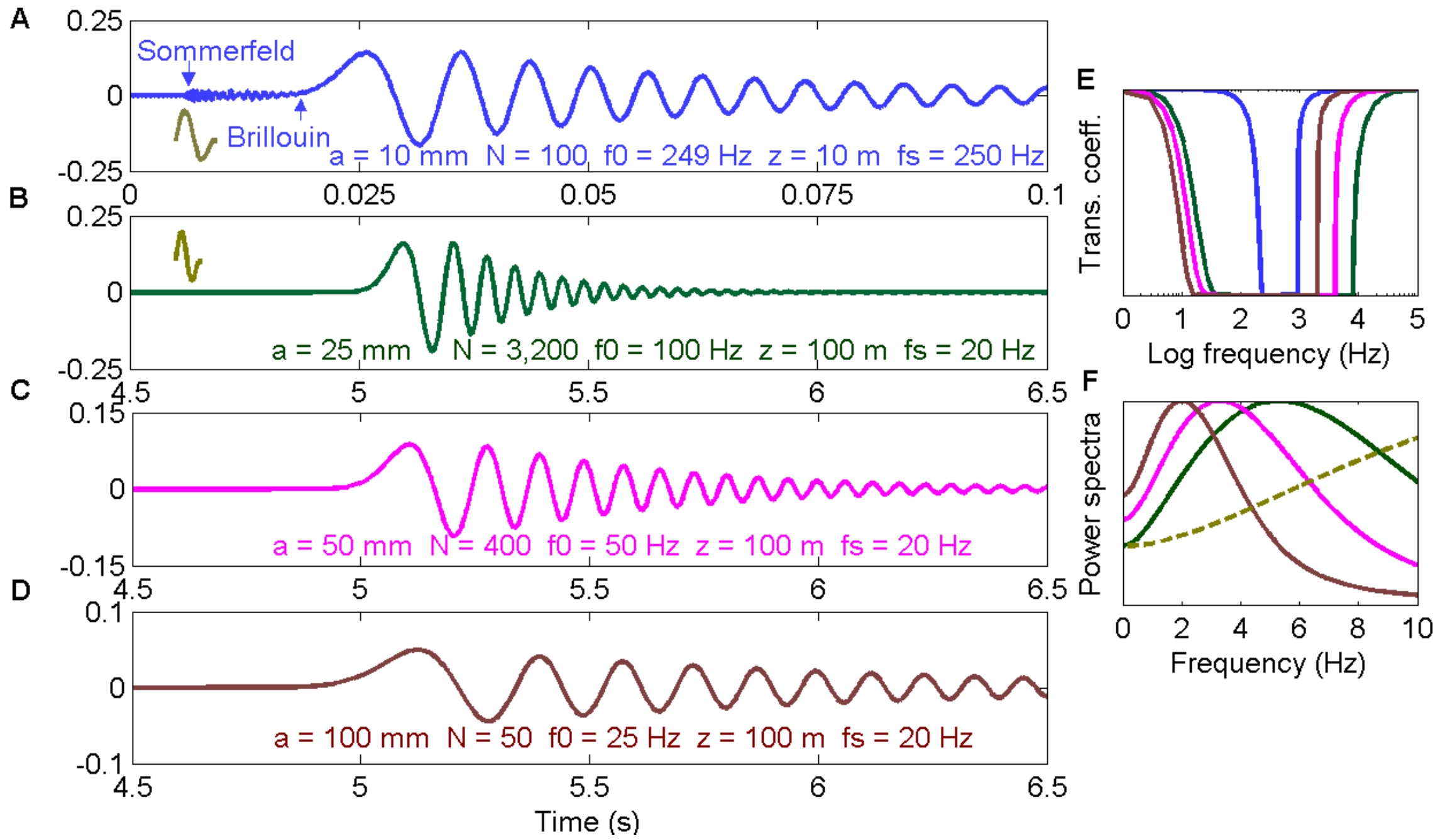

**Fig. 1.** Acoustic wave scattering by bubble cloud with different bubble radius. Incident wave is a single cycle pulse (solid olive, with scaled-down amplitude) with the dominant frequency $f_s = 250$ Hz or 20 Hz (dash olive). (**A**) $N$ = 100, $a$ = 10 mm, $z$ = 10 m, and $f_s = 250$ Hz (blue). (**B**) $N$ = 3,200 and $a$ = 25 mm (dark green). (**C**) $N$ = 400 and $a$ = 50 mm (magenta). (**D**) $N$ = 50 and $a$ = 100 mm (dark red). Propagation distance ($z$ = 100 m), bubble proportion ($\beta$ = 21%), and incident pulse ($f_s = 20$ Hz) are the same for (**B**), (**C**) and (**D**). (**E**) Transmission coefficients. (**F**) Normalized power spectra.

Figures 1B to 1D show the feature of Brillouin precursor field for different bubble radius but the same bubble proportion ($\beta$ = 21%) and propagation distance ($z$ = 100 m). The larger the bubble radius, the weaker the damping, and the lower the frequency of Brillouin precursor. The dominant frequencies of the first cycle Brillouin precursors in Fig. 1F are about 5.3 Hz for $a$ = 25 mm (dark green), about 3.3 Hz for $a$ = 50 mm (magenta), and about 2.0 Hz for $a$ = 100 mm (dark red). The spectra of Brillouin precursors are inversely proportional to the bubble radius and are about one order of magnitude lower (about 19, 15, and 13 times lower) than those of resonance of a single bubble.

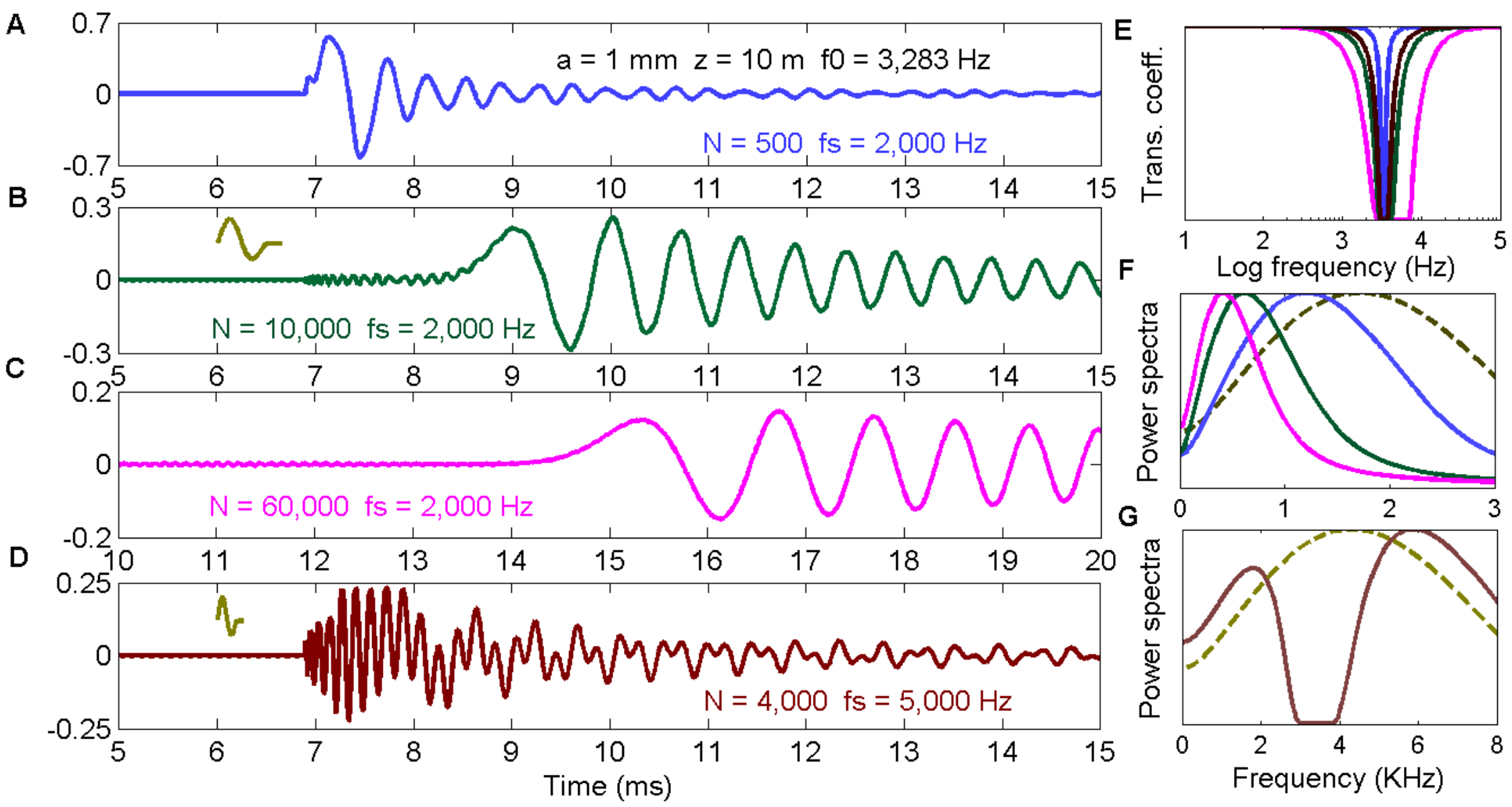

**Fig. 2.** Acoustic wave scattering by bubble cloud with different bubble proportion. Incident wave is a single cycle pulse (solid olive, with scaled-down amplitude) with dominant frequency $f_s = 2{,}000$ Hz or 5,000 Hz (dash olive). (**A**) $N$ = 500, $\beta = 0.0002\%$, and $f_s = 2{,}000$Hz (blue). (**B**) $N$ = 10,000, $\beta = 0.004\%$, and $f_s = 2{,}000$Hz (dark green). (**C**) $N$ = 60,000, $\beta = 0.03\%$, and $f_s = 2{,}000$Hz (magenta). (**D**) $N$ = 4000, $\beta = 0.002\%$, and $f_s = 5000$Hz (dark red). (**E**) Transmission coefficients. (**F, G**) Normalized power spectra.

Figures 2A to 2D show the acoustic scattering wave field for bubble cloud in water with the same bubble radius ($a$ = 1 mm) and propagation distance ($z$ = 10 m) but different bubble proportion. The other parameters are $\gamma = 1.4$, $\rho_f = 1{,}000$ kg/m$^3$, $v_f = 1{,}450$ m/s, $P_0 = 1.013 \times 10^5$ Pa, $b = 0.005\omega_0$. The features of the four calculated waveforms are similar in morphology to the four typical experiment waveforms classified from over 2000 cases of scattering of sound by bubble clouds (*14*). The most striking waveform features are a small saw-tooth wave for the early arrival in Fig. 2A and beating phenomenon for the wave packet evolution in Figs. 2D and 2G. The dominant frequencies of the first cycle Brillouin precursors are about 1,200 Hz for $\beta = 0.0002\%$, about 620 Hz for $\beta = 0.004\%$, and about 420 Hz for $\beta = 0.03\%$ (Fig. 2F). The large differences in the bubble proportion only produce small differences in the frequency. This indicates that bubble proportion has little influence on the frequency of Brillouin precursor field, however the bubble proportion has a significant influence on wave packet velocity, which decreases with increasing bubble proportion (Figs. 2A to 2C). This is because the effective velocity of bubble cloud medium ($v_e = \sqrt{K_e/\rho_e}$) is determined by the effective bulk modulus $K_e$ and density $\rho_e$; and a gas-bearing liquid medium has approximately the bulk modulus close to gas and the density close to liquid.

## Low-frequency Resonance in Strong 1D Heterogeneity

For a 1D heterogeneity, the delta matrix propagator approach (*10*) can provide an analytical solution that includes all multiple scattering effects (see Supplementary material), which may include more complex scattering phenomena than those of bubble cloud model. Two-constituent units embedded between two fluid half-spaces are used to simulate strong nonlinear interaction in 1D heterogeneity (*15 - 17*). The physical properties of constituent units are shown in Table 1. The strong acoustic impedance contrasts between the constituent units indicate that plastic/steel heterogeneity, shale/gas I heterogeneity, and shale/gas II heterogeneity are strong 1D heterogeneities. Different scale heterogeneities are constructed by varying the lattice constant *d* while the material proportions and the total thickness remain constants except Fig. 5. The incident pulse is a single cycle pulse (solid olive in Figs. 3 to 7, with scaled-down amplitude) with a dominant frequency of $f_s = 172$ Hz (dash olive in Figs. 3 to 7). This strong scattering problem belongs to the trapezoid-shaped region in $ka$ – $kL$ scattering classification diagram with 0.01 < *ka* < 10, where *a*, *k*, and *L* are the inhomogeneous scale length, the wavenumber, and the propagation distance, respectively (*9*).

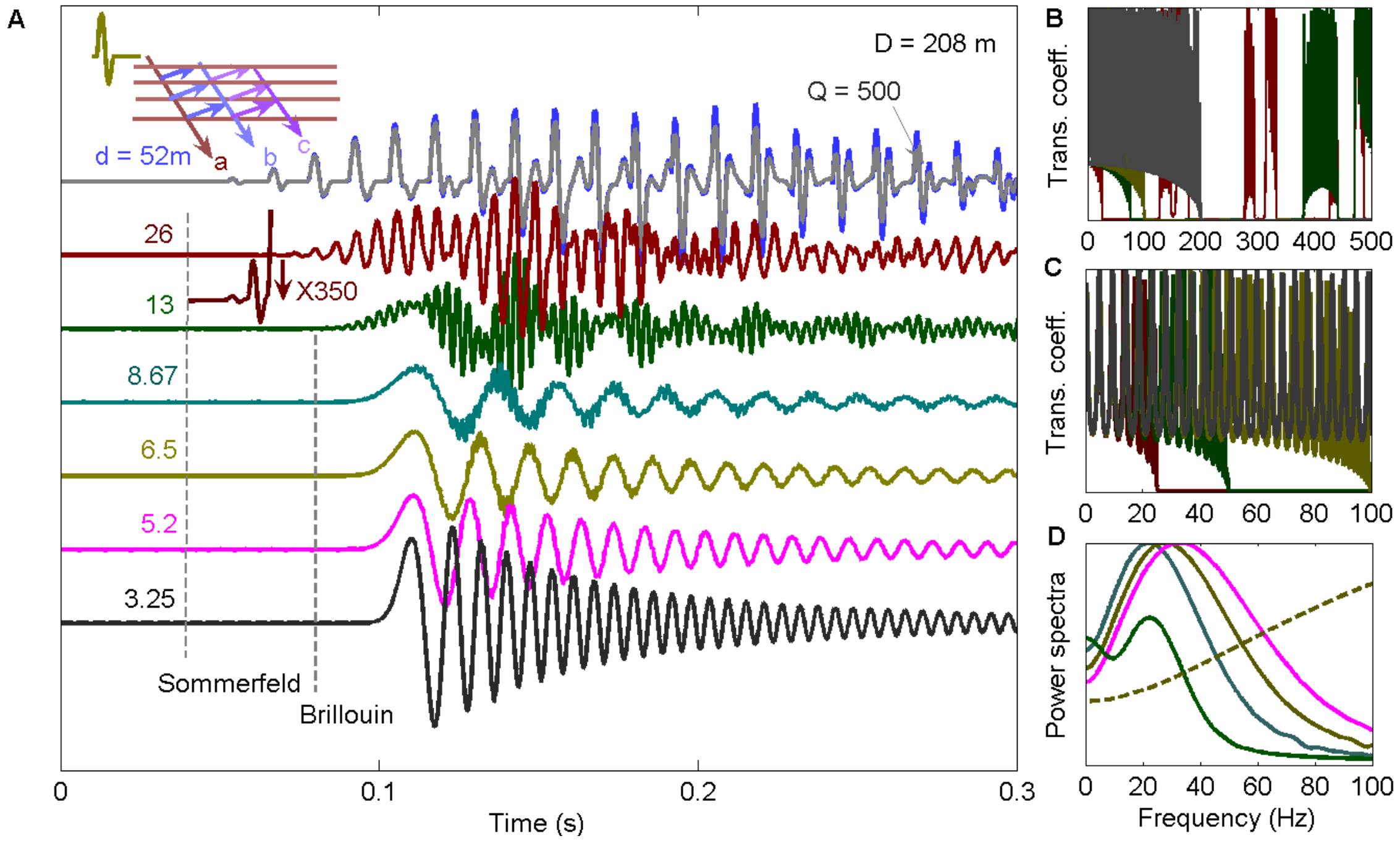

**Fig. 3.** Scale-dependent low-frequency resonance. Plastic/steel heterogeneity with a total thickness *D* = 208 m and different lattice constant *d* that varies from *d* = 52 m (8 layers, $d_1$ = 17 m, near seismic wavelength) to *d* = 3.25 m, (128 layers, $d_1$ = 1.0625 m, much less than seismic wavelength). (**A**) Normal transmission wave field. (**B, C**) Transmission coefficients. (**D**) Normalized power spectra of the first cyclic low-frequency resonance (LFR).

Figure 3 shows the normal transmission wave field, transmission coefficient, and normalized power spectrum for 1D plastic/steel heterogeneity with a total thickness $D = D_1 + D_2 = 208$ m (32.7% plastic with $D_1 = 68$ m and 67.3% steel with $D_2 = 140$ m) and different lattice constant *d* that varies from $d = d_1 + d_2 = 52$ m (plastic $d_1 = 17$ m and steel $d_2 = 35\,\mathrm{m}$) to $d$ = 3.25 m (plastic $d_1 = 1.0625\mathrm{m}$ and steel $d_2 = 2.1875\mathrm{m}$). The plastic thickness $d_1$ in plastic/steel heterogeneity, which is physically equivalent to the bubble radius $a$ in bubble cloud heterogeneity, can be seen as heterogeneous scale of medium if the steel is considered as background medium. The light grey for $d$ = 52 m stands for the medium with intrinsic absorption quality factor ($Q$ = 500), which only causes a slightly smaller amplitude than that of the corresponding non-absorption medium (blue). The influence of intrinsic absorption on wave packet evolution is weak and will be ignored in the following analysis.

The graphics at the top left of Fig. 3 depicts the direct and the multiple arrivals. The direct wave “a” has very small amplitude because of the transmission loss. The amplitudes from labels “a” to “b” to “c” etc. initially increase gradually and then decrease because of the constructive and destructive interferences of many multiple reflections. These amplitudes (or local extrema) form an upper or lower envelope with very low modulation frequency or low frequency background ($d$ = 52 m, $d_1$ = 17 m). This low-frequency background exhibits features of the Brillouin precursor field and the wavetrain “a”, “b”, “c”, etc. exhibits features of the Sommerfeld precursor field. As the lattice constant reduces ($d$ = 26 m, $d_1$ = 8.5 m), the amplitudes of the early arrivals (the direct wave and the follows) are very small and the very weak direct wave (the first arrival) is only visible by magnifying 350 times, thus the amplitude of the direct wave becomes negligible and the multiple waves become the first arrival (the behaviour of Sommerfeld precursor field). The corresponding low frequency background exhibits slightly more rapidly changing amplitude. As the lattice constant reduces further ($d$ = 13 m, $d_1$ = 4.25 m), the low frequency background gradually transfers into a real low-frequency component superposed on a high-frequency component (high-frequency onset). For smaller lattice constants ($d$ = 8.67 m or $d_1$ = 2.83 m to $d$ = 3.25 m or $d_1$ = 1.0625 m), the low-frequency component will transfer into a low-frequency primary with a very slowly rising edge. Its instantaneous frequency increases and its amplitude decreases with increasing propagation time. This wave packet evolution can be described by the hyper-Airy function (the behaviour of Brillouin precursor field). Finally the low-frequency wave will transfer into a direct transmission wave in an equivalent transversely isotropic medium for very small lattice constant $d \le 0.2\,\mathrm{m}$ (*17*).

The normalized power spectra of the first cyclic low-frequency component for different lattice constant ($d_1$ = 4.25 m, 2.83 m, 2.125 m, and 1.7 m) in Fig. 3A are shown in Fig. 3D. The dominant frequencies are about 22.5 Hz for $d_1$ = 2.83 m (dark cyan), about 27.5 Hz for $d_1$ = 2.125 m (dark olive green), and about 32.5 Hz for $d_1$ = 1.7 m (the magenta). Its frequencies are inversely proportional to the lattice constant or heterogeneous scale. The low-frequency component is due to the coherent scattering enhancement of multiple scattering waves in strong small-scale heterogeneity, which exhibits resonance-like wave phenomenon with high intensity and scale-dependent frequency. I call this phenomenon low-frequency resonance (LFR); a kind of collective behaviour or emergence phenomenon that occurs in transient regime. This modeling also shows that the plastic proportion has little influence on the frequency of LFR and exhibits the property similar to that of the bubble proportion in Fig. 2. Note that the concept of LFR is different from that of acoustic resonance scattering generated by the excitation of resonance or creeping wave of a single body during scattering process (*18*).

From the viewpoint of hierarchical structures, the scattering of wave field in Fig. 3A can be viewed as the superposition of the high-frequency and low-frequency wave components. The former is associated with Sommerfeld precursor and the latter Brillouin precursor in bubble cloud model. Sommerfeld precursor is predominant for large heterogeneous scale and Brillouin precursor for small heterogeneous scale. Sommerfeld precursor mainly exhibits the behaviour of individual constituents in the low hierarchy and Brillouin precursor as collective behaviour or emergence in the high hierarchy of the system. The scale-dependent transformation from the low to the high hierarchical structures is continuous because Sommerfeld and Brillouin precursors occurred in different hierarchical structures obey the same fundamental physics laws.

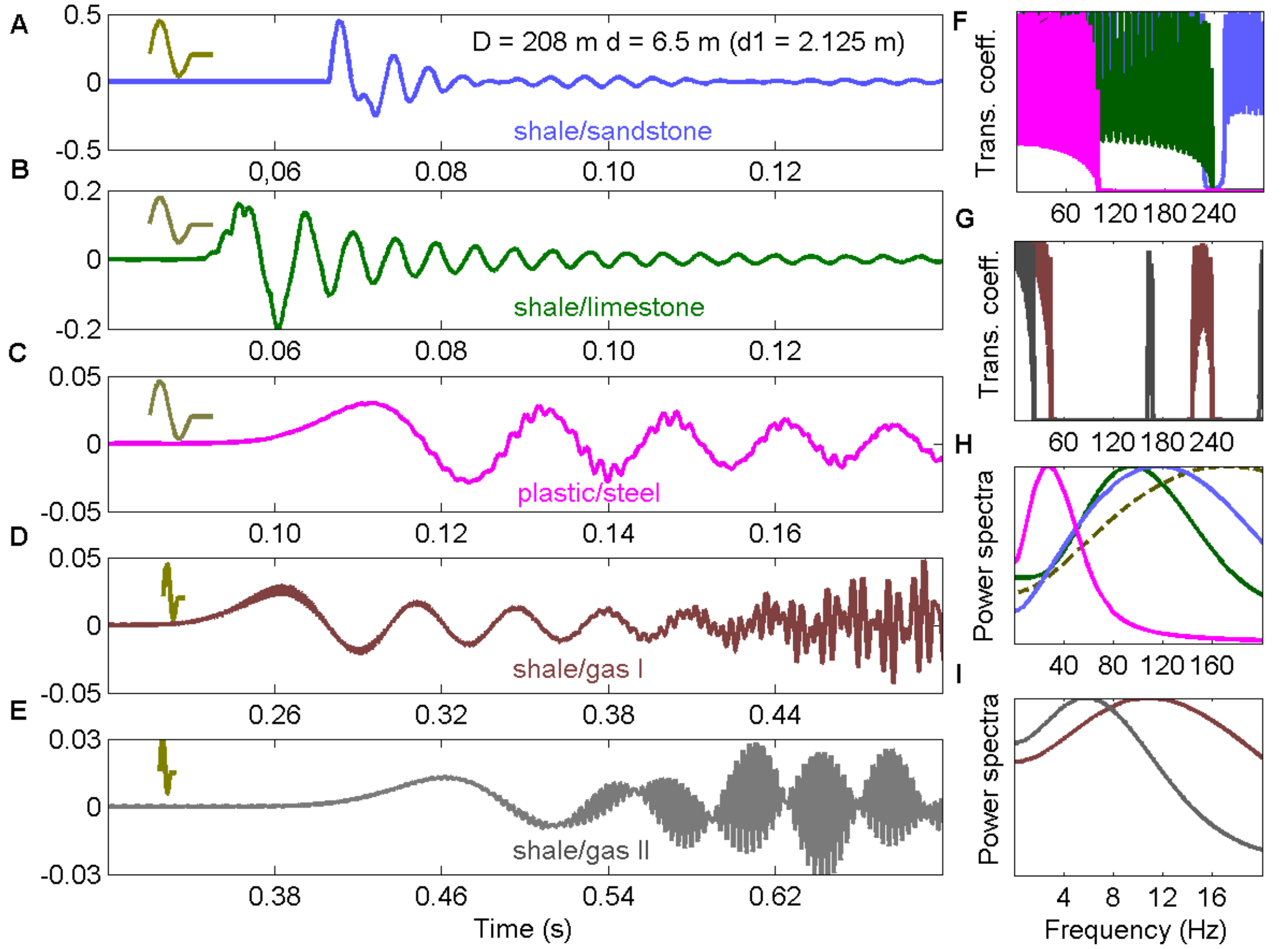

**Fig. 4.** Contrast-dependent low-frequency resonance. 1D heterogeneity with the same lattice constant *d* = 6.5 m ($d_1$ = 2.125 m) and total thickness *D* = 208 m and different constituents. (**A** – **E**) Normal transmission wave fields for shale/sandstone heterogeneity (blue), shale/limestone heterogeneity (dark green), plastic/steel heterogeneity (magenta), shale/gas I heterogeneity (dark red), and shale/gas II heterogeneity (grey). (**F, G**) Transmission coefficients. (**H, I**) Normalized power spectra.

Figure 4 shows the normal transmission wave field, transmission coefficient, and normalized power spectrum for 1D heterogeneities with the same lattice constant *d* = 6.5 m ($d_1$ = 2.125 m) and total thickness D = 208 m and five kinds of impedance contrasts. The larger the impedance contrast, the lower the frequency of the stopping band. This causes complex wave packet evolution in Figs. 4A to 4E. The frequencies of the first cyclic LFR are about 116 Hz for shale/sandstone heterogeneity, 95.5 Hz for shale/limestone heterogeneity, 27.5 Hz for plastic/steel heterogeneity, 11 Hz for shale/gas I heterogeneity, and 6 Hz for shale gas II heterogeneity. The frequency of LFR decreases with increasing impedance contrast of constituent units. The high-frequency small-amplitude saw-tooth waves superposing on the low-frequency background in Figs. 4C, 4D and 4E are associated with the resonances of individual constituent units. Their fundamental resonance frequency $f_0 = v_{pi}/2d_i$ (i = 1 and 2 stand for the constituent units) is 585 Hz for the plastic, 633 Hz for the steel, 235 Hz for the gas I, 165 Hz for the gas II, or 323 Hz for the shale. The fundamental resonance frequency of the individual plastic, gas I, or gas II is about 21 times for plastic/steel heterogeneity, 20 times for shale/gas I heterogeneity, or 27 times for shale gas II heterogeneity higher than the corresponding frequency of LFR.

Figure 5 shows the normal transmission wave field, transmission coefficient, and normalized power spectrum for plastic/steel heterogeneity with a lattice constant $d$ = 6.5 m ($d_1$ = 2.125 m) and four total medium thicknesses. The stopband corner frequencies are independent of the total thickness (Figs. 5B and 5C), however the rapid oscillations of transmission coefficient within the passbands are dependent on the total thickness; the thinner the thickness, the faster the oscillation. The frequencies of the first cyclic low-frequency resonance are about 27.5 Hz for D = 208 m, 24.5 Hz for D = 312 m, 22.5 Hz for D = 416 m, and 21 Hz for D =520 m. The frequency of LFR decreases marginally with increasing total medium thickness or propagation distance, and its amplitude also decreases marginally with the propagation distance. The longer the propagation distance, the smaller the relative changes of both the frequency and intensity of LFR. This indicates the low-frequency resonance is a kind of local resonance effect and is basically independent of the total medium thickness (or the total medium volume). This kind of localized wave, which is different from the classical Anderson's wave localization with very small diffusion constant or no diffusion (*19*), can be seen as a kind of "the ground state isolated at a lower energy" (*20*) and exhibits scattering propagation behaviour with no scattering attenuation or conduction-chain-like (*21*) and superconductivity-like (*22, 23*) and photonic-crystal-like (*24, 25*) propagation effects.

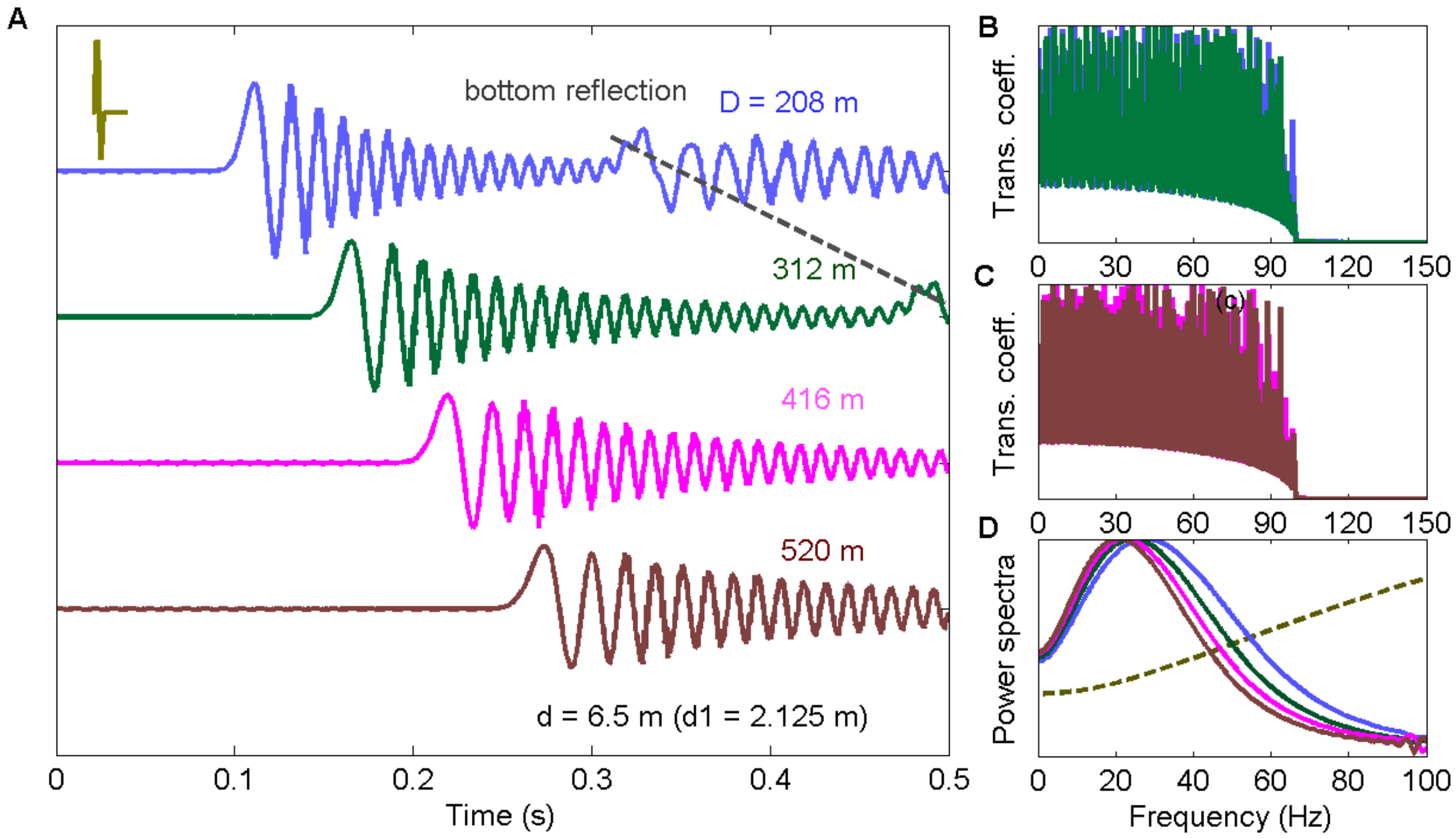

**Fig. 5.** Volume-independent low-frequency resonance. Plastic/steel heterogeneity with a lattice constant $d$ = 6.5 m ($d_1$ = 2.125 m) and four total thicknesses $D$ = 208 m (blue, 64 layers), $D$ = 312 m (dark green, 96 layers), $D$ = 416 m (magenta, 128 layers), and $D$ = 520 m (dark red, 160 layers). The straight dash grey denotes the reflections from the bottom fluid half-space. (**A**) Normal transmission wave fields. (**B, C**) Transmission coefficients. (**D**) Normalized power spectra.

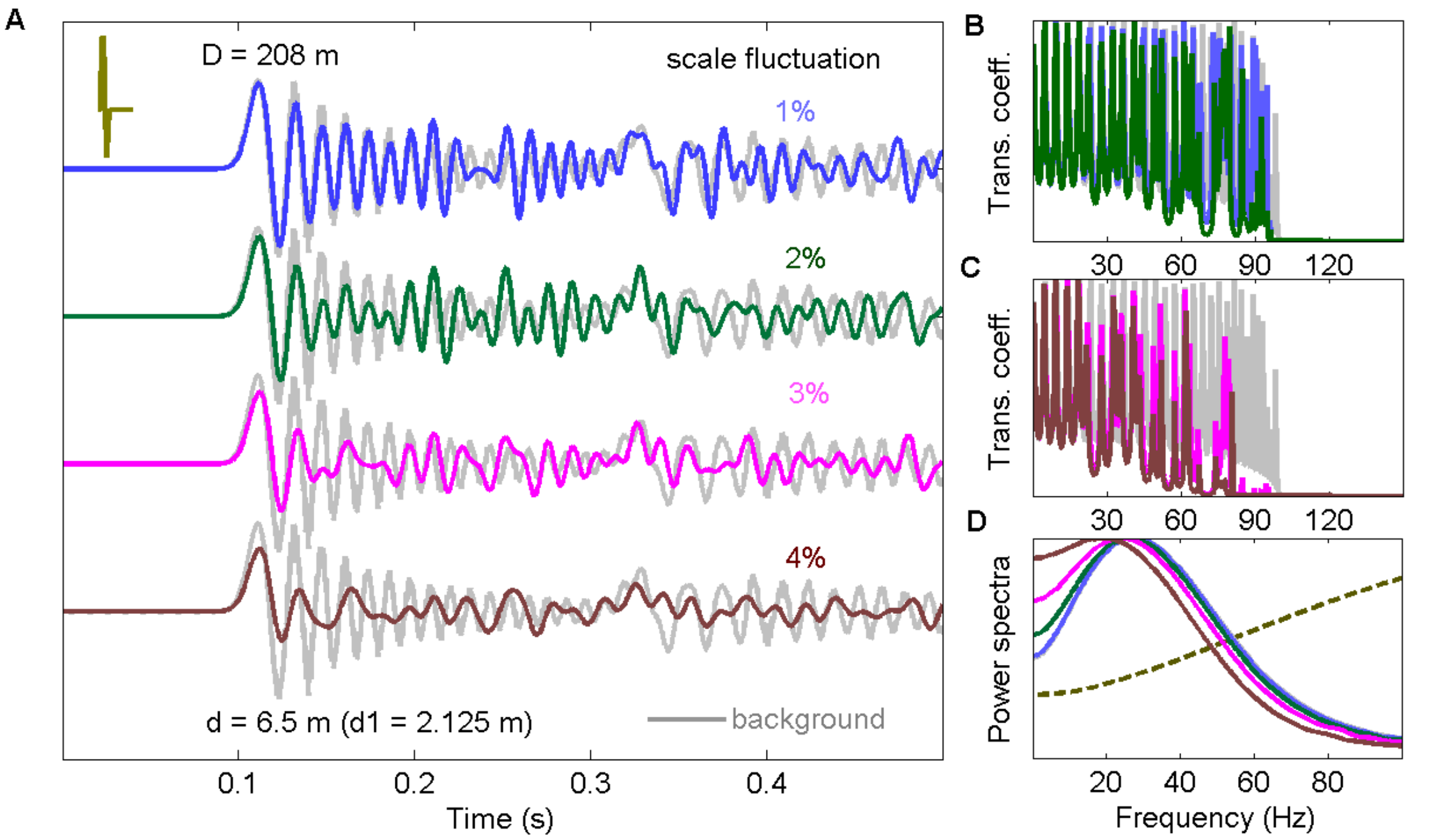

**Fig. 6.** Effect of random scale fluctuation on low-frequency resonance. Plastic/steel heterogeneity with lattice constant $d$ = 6.5 m ($d_1$ = 2.125 m), total thickness $D$ = 208 m,

and different RMS scale fluctuations. (**A**) Normal transmission wave fields for the scale fluctuations $\delta d/d$ = 1% (blue), 2% (dark green), 3% (magenta), and 4% (dark red). (**B, C**) Transmission coefficients. (**D**) Normalized power spectra.

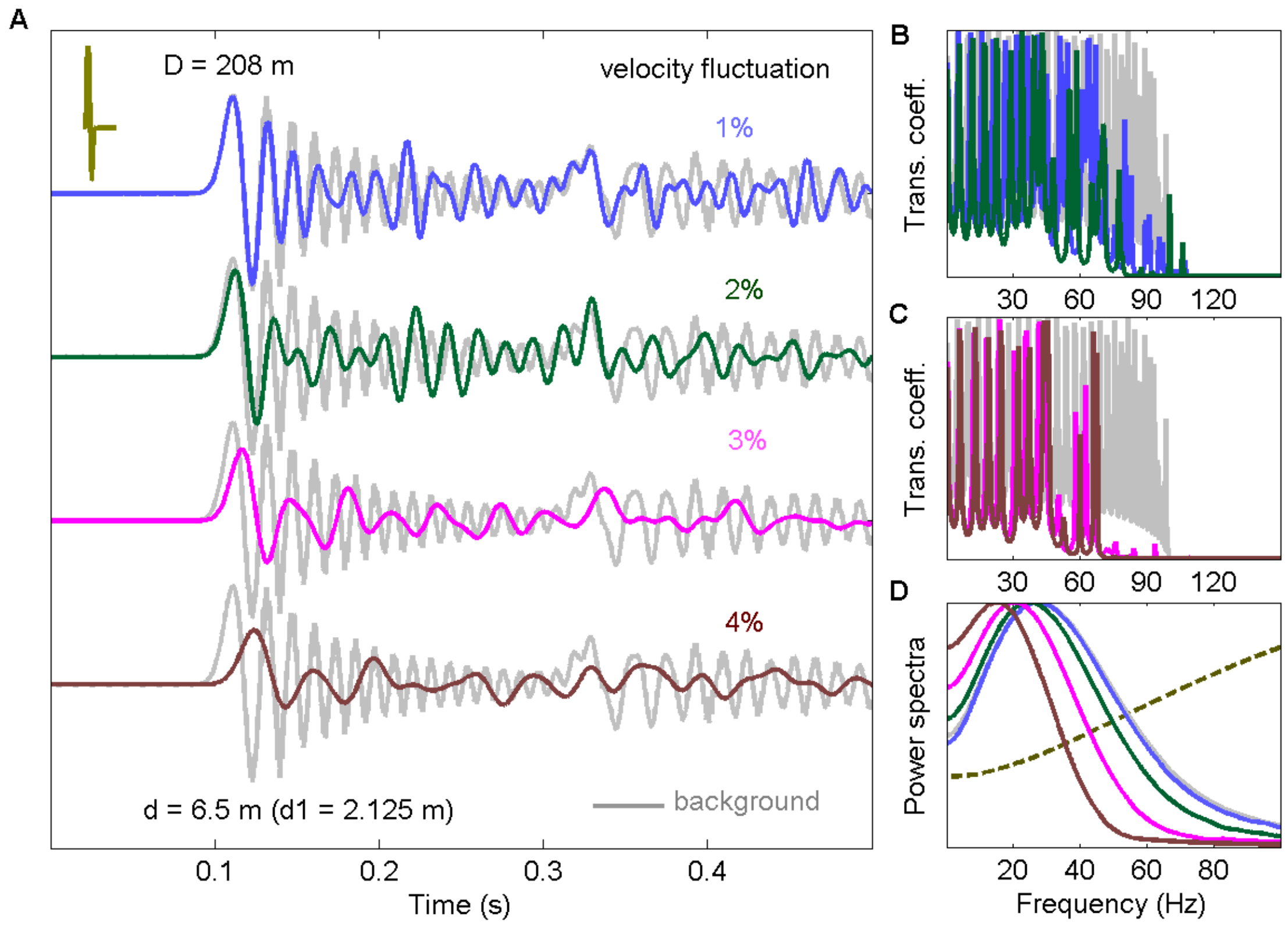

**Fig. 7.** Effect of random velocity fluctuation on low-frequency resonance. The same as Fig. 6 except for RMS velocity fluctuations $\delta v/v$ = 1% (blue), 2% (dark green), 3% (magenta), and 4% (dark red).

Figures 6 and 7 show the influence of random scale (Fig. 6) and velocity (Fig. 7) fluctuations of plastic/steel heterogeneity on low-frequency resonance. The fluctuations labeled from 1% to 4% (Figs. 6A and 7A) represent the root-mean-square (RMS) scale and velocity fluctuations (the grey for the background), respectively. An increase in the RMS scale and velocity fluctuations means a decrease in the spatial symmetry of small-scale heterogeneity. It can be seen in Figs. 6B, 6C, 7B and 7C that the first stopband corner frequency shifts slightly toward lower frequency and the oscillation peaks decrease slightly with the increasing RMS scale and velocity fluctuations (the grey for the background). The frequencies of the first cyclic low-frequency resonance are about 27.5 Hz for $\delta d/d$ = 0% (grey) and 1% (blue), 26.5 Hz for $\delta d/d$ = 2% (dark green), 24 Hz for $\delta d/d$ = 3% (magenta), and 19 Hz for $\delta d/d$ = 4% (dark red) for scale fluctuations in Fig. 6D; and are about 27.5 Hz for $\delta v/v$ = 0% (grey) and 1% (blue), 25 Hz for $\delta v/v$ = 2% (dark green), 20.5 Hz for $\delta v/v$ = 3% (magenta), and 15 Hz for $\delta v/v$ = 4% (dark red) for velocity fluctuations in Fig. 7D. LFR is a little more sensitive to the velocity than the scale fluctuations. The frequency of LFR decreases with increasing random heterogeneous scale and velocity fluctuations; and its energy also decreases with increasing scale and velocity fluctuations (Figs. 6A and 7A). This feature suggests that the frequency and strength of LFR will decrease with lowering degree of spatial symmetry of small-scale heterogeneity.

## Discussion

Observed low-frequency seismic anomalies or long period events are always associated with strong small-scale seismic heterogeneity. For example, hydraulic fracturing microseismicity (*3*), volcanic tremor (*1*, *2*), and non-volcanic tremor (*26*, *27*). LFR provides a physical interpretation for the observed low-frequency phenomena. Low-frequency resonance originates from the interference or coherence of multiple scattering waves and should be a ubiquitous phenomenon in wave physics. It is believed that the observed low-frequency anomalies or the coexistence of high-frequency and low-frequency oscillations in wave-particle interactions, that include electromagnetic, matter, and gravitational waves (*4 - 7*, *20 - 25*, *28*, *29)*, are also associated with LFR.

LFR is a kind of collective behaviour or emergence phenomenon caused by multiple wave scattering in strong small-scale heterogeneity. Emergence phenomenon is the origin of many fascinating phenomena in nature with scales ranging from the smallest subatomic particles to the largest universe stars. The classic multiple scattering theory (MST) provides exact analytical series solutions for 2D and 3D many-body systems (*30*). These solutions can be developed to numerically study the microscopic constituent interactions and the macroscopic collective behaviour in more complex 2D and 3D many-body systems. Random matrix theory (RMT) studies the eigenvalue spacing distribution of response matrix for evaluating the symmetries and collectivities of the microscopic constituent units (*31*). The marriage between MST and RMT may develop the technologies with subwavelength spatial resolution for understating the microscopic constituent distribution of a complex many-body system.

## Tables

**Table 1** Physical properties of constituents

| Medium | $v_p(m/s)$ | $v_s(m/s)$ | $\rho(kg/m^3)$ |
|---|---|---|---|
| Plastic | 2487 | 1048 | 1210 |
| Steel | 5535 | 3000 | 7900 |
| Shale | 2743 | 1509 | 2380 |
| Sandstone | 3353 | 1844 | 2300 |
| Limestone | 5540 | 3040 | 2700 |
| Gas I | 1000 | | 400 |
| Gas II | 700 | | 250 |

**Acknowledgements**

I thank my wife, Xiaoping Dai and my daughter, Wenbo Elissa Liu for their encouragement and understanding that keep my inner stability for the past over ten years. I thank Professor Erik Eberhardt of The University of British Columbia for providing me a visiting opportunity to revise the manuscript.

**Data and Resources Section**

No data were used in this paper.

**Supplemental Material:** This paper includes electronic supplemental materials that describe the methods for simulating the scattering wave fields in bubble cloud heterogeneity and in 1D heterogeneity.

## Methods

### Bubble Cloud Model

For acoustic wave scattering in bubble cloud medium, based on Foldy's multiple scattering theory, the effective wavenumber (*13*) and the acoustic wave field in time domain can be written as

$$k^2 = \frac{\omega^2}{v_f^2}\left(1 + \frac{4\pi v_f^2 N a}{\omega^2 + 2i\omega b - \omega_0^2}\right) \quad \text{(S1)}$$

$$\omega_0 = 2\pi f_0 = \frac{1}{a}\sqrt{\frac{3\gamma P_0}{\rho_f}} \quad \text{(S2)}$$

$$p(z,t) = \frac{1}{2\pi}\mathrm{Re}\int G(\omega)\exp[i(kz - \omega t)]d\omega \quad \text{(S3)}$$

where Re means "the real part", $k$ is the effective wavenumber, and $G(\omega)$ is the spectrum of a plane incident pulse. $N$, $a$, $\omega_0$, $b$, $\gamma$, $\rho_f$, $v_f$, and $P_0$ are the number of bubbles per unit volume, the radius of the bubble, the Minnaert resonance angular frequency (*12*), damping constant, the ratio of specific heats, the density, the acoustic velocity, and the hydrostatic pressure, respectively.

The transmission coefficients and wave fields for bubble cloud scattering can be calculated by equations (S1) to (S3).

### Delta-Matrix Propagator Approach

Propagator matrix approach (*9*) can provide an exact analytical solution for 1D heterogeneity, however there is computational instability for the reflection and transmission coefficients. Delta matrix propagator (*10*) can improve the computational instability and is employed to study the multiple scattering processing in this study.

The displacement and stress matrix can be written as

$$S_0^- = B S_n^+ \quad \text{(S4)}$$

$$B = \prod_{i=1}^{n} B_i \quad \text{(S5)}$$

$$B_i = X_i D_i^{-1} X_i^{-1} \quad \text{(S6)}$$

where $S = (u_x, u_z, \sigma_{zz}, \sigma_{zx})$ is the displacement and stress vector. $X_i$, $D_i$, and $B_i$ are $4\times 4$ matrixes related to medium properties.

$$X_i = \begin{pmatrix} 1 & 1 & 1 & 1 \\ \xi_{1i} & -1/\xi_{2i} & -\xi_{1i} & 1/\xi_{2i} \\ \rho_i(v^2 - 2\beta_i^2) & -2\rho_i\beta_i^2 & \rho_i(v^2 - 2\beta_i^2) & -2\rho_i\beta_i^2 \\ 2\rho_i\beta_i^2\xi_{1i} & \rho_i(v^2 - 2\beta_i^2)/\zeta_{2i} & -2\rho_i\beta_i^2\xi_{1i} & -\rho_i(v^2 - 2\beta_i^2)/\xi_{2i} \end{pmatrix} \quad \text{(S7)}$$

$$D_i = \begin{pmatrix} \exp(ix_{1i}) & 0 & 0 & 0 \\ 0 & \exp(ix_{2i}) & 0 & 0 \\ 0 & 0 & \exp(-ix_{1i}) & 0 \\ 0 & 0 & 0 & \exp(-ix_{2i}) \end{pmatrix} \quad \text{(S8)}$$

$$x_{1i} = \omega/v\xi_{1i}d_i \quad \text{(S9)}$$

$$x_{2i} = \omega/v\xi_{2i}d_i \quad \text{(S10)}$$

$$\xi_{1i} = \sqrt{v^2/\alpha_i^2 - 1} \quad \text{(S11)}$$

$$\xi_{2i} = \sqrt{v^2/\beta_i^2 - 1} \quad \text{(S12)}$$

where $\alpha_i$, $\beta_i$, and $d_i$ are the compressional and shear velocities and the thickness of layer i, respectively.

The reflection and transmission coefficients can be written as

$$R(\omega) = (R_1 - R_2)/(R_1 + R_2) \quad \text{(S13)}$$

$$T(\omega) = -2abb_{41}/(R_1 + R_2) \quad \text{(S14)}$$

$$R_1 = a[b(b_{31}b_{43} - b_{33}b_{41}) + a(b_{31}b_{42} - b_{32}b_{41})] \quad \text{(S15)}$$

$$R_2 = b[a(b_{21}b_{42} - b_{22}b_{41}) + b(b_{21}b_{43} - b_{23}b_{41})] \quad \text{(S16)}$$

$$a = \sqrt{v^2/v_f^2 - 1} \quad \text{(S17)}$$

$$b = \rho_f v^2 \quad \text{(S18)}$$

where $\rho_f$ and $v_f$ are the density and velocity of the fluid and v is the phase velocity. The transmission and reflection wave fields for an incident plane pulse with spectrum $G(\omega)$ can be written as

$$p_r(t) = \int_{-\infty}^{\infty} G(\omega)R(\omega)\exp[i(k\,x - \omega\,t)]d\omega \quad \text{(S19)}$$

$$p_t(t) = \int_{-\infty}^{\infty} G(\omega)T(\omega)\exp[i(k\,x - \omega\,t)]d\,\omega \quad \text{(S20)}$$

There is inherent computational instability in equations (S19) and (S20). The delta matrix propagator (*10*) can provide an analytical solution that accurately includes all propagation and scattering effects like multiple scattering, conversion of P and SV waves, and evanescence waves, *et al.*. The 2th-order delta subdeterminants of propagator B in equation (S5) can be written as

$$B_{IJ}^{\Delta} = b_{kl}^{ij} = b_{ik}b_{jl} - b_{il}b_{jk} \quad \text{(S21)}$$

where I and J = 1, 2, 3, 4, 5, 6 are corresponding to the paired indices ij or kl = 12, 13, 14, 23, 24, 34, respectively. Thus equations (S15) and (S16) can be expressed by delta matrix as

$$R_1 = a(bb_{62}^{\Delta} + ab_{61}^{\Delta}) \quad \text{(S22)}$$

$$R_2 = b(ab_{51}^{\Delta} + bb_{52}^{\Delta}) \quad \text{(S23)}$$

The elements of propagator matrix B are

$$b_{11} = b_{44} = \frac{2\beta_i^2 \cos x_{1i} + (v^2 - 2\beta^2)\cos x_{2i}}{v^2} \quad \text{(S24)}$$

$$b_{12} = b_{34} = \frac{i\left[(2\beta_i^2 - v^2)\sin x_{1i} + 2\beta_i^2 \xi_{1i}\xi_{2i})\sin x_{2i}\right]}{\xi_{1i} v^2} \quad \text{(S25)}$$

$$b_{13} = b_{24} = \frac{\cos x_{1i} - \cos x_{2i}}{\rho_i v^2} \quad \text{(S26)}$$

$$b_{14} = -i\frac{\sin x_{1i} + \xi_{1i}\xi_{2i}\sin x_{2i}}{\rho_i v^2 \xi_{1i}} \quad \text{(S27)}$$

$$b_{21} = b_{43} = i\frac{-2\xi_{1i}\xi_{2i}\beta_i^2 \sin x_{1i} + (v^2 - 2\beta_i^2)\sin x_{2i}}{\xi_{2i} v^2} \quad \text{(S28)}$$

$$b_{22} = b_{33} = \frac{(v^2 - 2\beta_i^2)\cos x_{1i} + 2\beta_i^2 \cos x_{2i}}{v^2} \quad \text{(S29)}$$

$$b_{23} = -i\frac{\xi_{1i}\xi_{2i}\sin x_{1i} + \sin x_{2i}}{\rho_i v^2 \xi_{2i}} \quad \text{(S30)}$$

$$b_{31} = b_{42} = \frac{2\rho_i\beta_i^2 (v^2 - 2\beta_i^2)(\cos x_{1i} - \cos x_{2i})}{v^2} \quad \text{(S31)}$$

$$b_{32} = -i\frac{\rho_i (v^2 - 2\beta_i^2)^2 \sin x_{1i} + 4\rho_i\beta_i^4 \xi_{1i}\xi_{2i}\sin x_{2i}}{v^2 \xi_{1i}} \quad \text{(S32)}$$

$$b_{41} = -i\frac{4\rho_i\beta_i^4 \xi_{1i}\xi_{2i}\sin x_{1i} + \rho_i (v^2 - 2\beta_i^2)^2 \sin x_{2i}}{v^2 \xi_{2i}} \quad \text{(S33)}$$

The elements of delta propagator $B^{\Delta}$ are

$$b_{11}^{\Delta} = b_{66}^{\Delta} = \frac{1}{v^4 \xi_{1i}\xi_{2i}}[4\zeta_{1i}\zeta_{2i}\beta_i^2 (2\beta_i^2 - v^2)(\cos x_{1i}\cos x_{2i} - 1)$$
$$-(4\beta_i^4 - 4\beta_i^2 v^2 + v^4 + 4\xi_{1i}^2\xi_{2i}^2\beta_i^4)\sin x_{1i}\sin x_{2i}] + 1 \quad \text{(S34)}$$

$$b_{12}^{\Delta} = b_{56}^{\Delta} = -\frac{i}{\rho_i v^2}\left(1/\xi_{2i} \cos x_{1i}\sin x_{2i} + \xi_{1i}\sin x_{1i}\cos x_{2i}\right) \quad \text{(S35)}$$

$$b_{13}^{\Delta} = b_{14}^{\Delta} = b_{36}^{\Delta} = b_{46}^{\Delta} = \frac{1}{\rho_i v^4 \xi_{1i}\xi_{2i}}[\zeta_{1i}\zeta_{2i}(v^2 - 4\beta_i^2)(\cos x_{1i}\cos x_{2i} - 1)$$
$$+(2\beta_i^2 - v^2 + 2\xi_{1i}^2\xi_{2i}^2\beta_i^2)\sin x_{1i}\sin x_{2i}] \quad \text{(S36)}$$

$$b_{15}^{\Delta} = b_{26}^{\Delta} = -i\frac{1}{\rho_i v^2}\left(1/\xi_{1i} \sin x_{1i}\cos x_{2i} + \xi_{2i}\sin x_{2i}\cos x_{1i}\right) \quad \text{(S37)}$$

$$b_{16}^{\Delta} = \frac{1}{\rho_i^2 v^4}\left\{-2\cos x_{1i}\cos x_{2i} + \left[\xi_{1i}\xi_{2i} + 1/(\xi_{1i}\xi_{2i})\right]\sin x_{1i}\sin x_{2i} + 2\right\} \quad \text{(S38)}$$

$$b_{21}^{\Delta} = b_{65}^{\Delta} = -\frac{i\rho_i}{v^2}\left[4\xi_{2i}\beta_i^4 \cos x_{1i}\sin x_{2i} + 1/\xi_{1i}\left(4\beta_i^4 - 4\beta_i^2 v^2 + v^4\right)\sin x_{1i}\cos x_{2i}\right] \quad \text{(S39)}$$

$$b_{22}^{\Delta} = b_{55}^{\Delta} = \cos x_{1i}\cos x_{2i} \quad \text{(S40)}$$

$$b_{23}^{\Delta}=b_{24}^{\Delta}=b_{35}^{\Delta}=b_{45}^{\Delta}=\frac{i}{v^2}\left[2\beta_i^2\xi_{2i}\cos x_{1i}\sin x_{2i}+(2\beta_i^2-v^2)/\xi_{1i}\sin x_{1i}\cos x_{2i}\right] \tag{S41}$$

$$b_{25}^{\Delta}=\xi_{2i}/\xi_{1i}\sin x_{1i}\sin x_{2i} \tag{S42}$$

$$\begin{aligned}b_{31}^{\Delta}=b_{64}^{\Delta}=b_{41}^{\Delta}=b_{63}^{\Delta}=\frac{\rho_i}{v^4\xi_{1i}\xi_{2i}}&[2\zeta_{1i}\zeta_{2i}\beta_i^2(8\beta_i^4-6\beta_i^2v^2+v^4)(\cos x_{1i}\cos x_{2i}-1)\\ &-\left(6\beta_i^2v^4+8\beta_i^6-12\beta_i^4v^2-v^6+8\xi_{1i}^2\xi_{2i}^2\beta_i^6\right)\sin x_{1i}\sin x_{2i}]\end{aligned} \tag{S43}$$

$$b_{32}^{\Delta}=b_{42}^{\Delta}=b_{53}^{\Delta}=b_{54}^{\Delta}=-\frac{i}{v^2}\left[2\beta_i^2\xi_{1i}\sin x_{1i}\cos x_{2i}+(2\beta_i^2-v^2)/\xi_{2i}\sin x_{2i}\cos x_{1i}\right] \tag{S44}$$

$$\begin{aligned}b_{33}^{\Delta}=b_{44}^{\Delta}=\frac{1}{v^4\xi_{1i}\xi_{2i}}&[4\zeta_{1i}\zeta_{2i}\beta_i^2(v^2-2\beta_i^2)(\cos x_{1i}\cos x_{2i}-1)\\ &+\left(4\beta_i^4+v^4+4\xi_{1i}^2\xi_{2i}^2\beta_i^4+4\beta_i^2v^2\right)\sin x_{1i}\sin x_{2i}]+1\end{aligned} \tag{S45}$$

$$b_{34}^{\Delta}=b_{43}^{\Delta}=b_{33}^{\Delta}-1 \tag{S46}$$

$$b_{51}^{\Delta}=b_{62}^{\Delta}=\frac{i\rho_i}{v^2\xi_{2i}}\left[\left(4\beta_i^4-4\beta_i^2v^2+v^4\right)\sin x_{2i}\cos x_{1i}+4\beta_i^4\xi_{1i}\xi_{2i}\sin x_{1i}\cos x_{2i}\right] \tag{S47}$$

$$b_{52}^{\Delta}=\xi_{1i}/\xi_{2i}\sin x_{1i}\sin x_{2i} \tag{S48}$$

$$\begin{aligned}b_{61}^{\Delta}=-\frac{\rho_i^2}{v^4\xi_{1i}\xi_{2i}}&[8\zeta_{1i}\zeta_{2i}\beta_i^4(4\beta_i^4-4\beta_i^2v^2+v^4)(\cos x_{1i}\cos x_{2i}-1)\\ &-\left(16\beta_i^8-32\beta_i^6v^2+24\beta_i^4v^4-8\beta_i^2v^6+v^8+16\xi_{1i}^2\xi_{2i}^2\beta_i^8\right)\sin x_{1i}\sin x_{2i}]\end{aligned} \tag{S49}$$

Thus, the reflection and transmission coefficients in equations (S13) and (S14) and the reflection and transmission wave fields in equations (S19) and (S20) can be calculated by using equations (S22) and (S23) as well as equations (S24) to (S49).